\documentclass[aps, prx, superscriptaddress, reprint, floatfix]{revtex4-2}
\usepackage{graphicx} 
\usepackage{amsmath}
\usepackage{amsfonts}
\usepackage{braket}
\usepackage{tikz}
\usepackage{subcaption}
\usepackage{todonotes}
\usepackage{algorithm}
\usepackage{algpseudocode}
\usepackage{multirow}
\usepackage{rotating}
\usepackage{tikz}
\usepackage[compat=0.6]{yquant}
\useyquantlanguage{groups}
\usepackage{comment}
\usepackage{xcolor}
\usepackage{gensymb}
\usepackage[colorlinks = true, linkcolor=blue, citecolor=blue, urlcolor  = blue]{hyperref}
\hypersetup{anchorcolor = black}
\usepackage{orcidlink}
\usepackage{times}

\def\equationautorefname~#1\null{Eq.~(#1)\null}

\begin{document}
\title{Equivalence checking of quantum circuits \\ via intermediary matrix product operator}

\author{Aaron Sander\,\orcidlink{0009-0007-9166-6113}}
\email{aaron.sander@tum.de}
\affiliation{
Chair for Design Automation, Technical University of Munich, Munich, Germany
}

\author{Lukas Burgholzer\,\orcidlink{0000-0003-4699-1316}}
\email{lukas.burgholzer@tum.de}
\affiliation{
Chair for Design Automation, Technical University of Munich, Munich, Germany
}
\affiliation{Munich Quantum Software Company GmbH, Garching near Munich, Germany}

\author{Robert Wille\,\orcidlink{0000-0002-4993-7860}}
\email{robert.wille@tum.de}
\affiliation{
Chair for Design Automation, Technical University of Munich, Munich, Germany
}
\affiliation{Munich Quantum Software Company GmbH, Garching near Munich, Germany}
\affiliation{Software Competence Center Hagenberg GmbH (SCCH), Hagenberg, Austria}

\date{\today{}}

\begin{abstract}
As quantum computing advances, the complexity of quantum circuits is rapidly increasing, driving the need for robust methods to aid in their design. Equivalence checking plays a vital role in identifying errors that may arise during compilation and optimization of these circuits and is a critical step in quantum circuit verification.
In this work, we introduce a novel method based on matrix product operators (MPOs) for determining the equivalence of quantum circuits. Our approach contracts tensorized quantum gates from two circuits into an intermediary MPO, exploiting their reversibility to determine their equivalence or non-equivalence. Our results show that this method offers significant scalability improvements over existing methods, with polynomial scaling in circuit width and depth for the practical use cases we explore. We expect that this work sets the new standard for scalable equivalence checking of quantum circuits and will become a crucial tool for the validation of increasingly complex quantum systems.
\end{abstract}
\maketitle

\section{Introduction}
Real quantum devices are constrained by limitations in gate sets, topologies, and noise characteristics, making the direct execution of ideal quantum algorithms impractical. Consequently, algorithms must undergo multiple layers of transformation, including compilation to available gate sets, mapping onto the architecture's topology, and optimization to minimize circuit depth and qubit usage. Each stage in this process introduces the potential for errors, leading to circuits that may fail to correctly represent the intended algorithm \cite{paltenghi_bugs_2022}. As quantum circuits grow in size, particularly with the advent of fault-tolerant quantum computing, these challenges will be further compounded by increased circuit depth and complexity. Without robust methods to identify and mitigate such errors, we risk a "design gap"—where quantum hardware advances, but the ability to reliably develop large-scale, executable algorithms lags behind. This underscores the pressing need for effective debugging tools to ensure that quantum circuits meet their intended functionality at each step of the development process~\cite{di_matteo_need_2024}.

Fortunately, the development of classical computing provides valuable lessons that can be leveraged to advance the design, verification, and debugging of quantum algorithms. In classical computing, formal equivalence checking has become an essential tool in electronic design automation, used to ensure that digital integrated circuits represent the same logical functionality, i.e., the same truth table \cite{drechsler_advanced_2004}. Drawing inspiration from this, we turn our focus to quantum circuits, aiming to verify whether two circuits implement the same unitary operation—a process known as equivalence checking.

Determining whether two quantum circuits are equivalent is a QMA-complete problem \cite{janzing_identity_2003}, requiring the development of advanced methods that not only operate efficiently in sub-exponential time but also scale effectively as quantum algorithms grow in complexity. Previous approaches to equivalence checking, such as those using ZX-Calculus, decision diagrams (DDs), or tensor decision diagrams (TDDs), face significant challenges in scalability. In this work, we introduce a novel equivalence checking method based on tensor networks—specifically, tensorized quantum circuits and matrix product operators (MPOs), a leading framework for simulating quantum many-body systems \cite{pirvu_matrix_2010, hubig_generic_2017, parker_local_2020}. Tensor networks offer a promising path to enhance the scalability of equivalence checking, allowing for error detection in larger circuits and supporting the development of more complex quantum algorithms. Furthermore, this approach leverages decades of established techniques in tensor networks, bringing a wealth of theoretical and practical knowledge to the equivalence checking problem.

To check the equivalence of two quantum circuits, we utilize an intermediary matrix product operator (MPO) to compare tensorized versions of the circuits. Given two circuits, \( G \) and \( G' \), the equivalence can be determined by computing \( GG'^\dagger \) and comparing it to the \(n\)-qubit identity operator, \( I_n \). If \( GG'^\dagger \) is approximately equal to \( I_n \) (within a specified precision), the circuits are equivalent; otherwise, they are non-equivalent.

In our proposed method, an MPO representing \( I_n \) is positioned between the two circuits, and gates from \( G \) and \( G'^\dagger \) are contracted into the MPO using an alternating scheme, \( G \rightarrow I_n \leftarrow G'^\dagger \), to exploit the desired equivalence property. This process resembles a double-sided time-evolving block decimation (TEBD) algorithm \cite{vidal_efficient_2003, bridgeman_hand-waving_2017, paeckel_time-evolution_2019}, where gates are applied in a two-qubit temporal zone to neighboring qubit pairs, enhancing computational efficiency and minimizing the growth of bond dimensions—especially when the circuits are equivalent or near-equivalent. Once all gates are contracted into the intermediary MPO, the equivalence of \( G \) and \( G' \) is evaluated by calculating the Frobenius norm of the resulting MPO \( GG'^\dagger \), based on its trace.

Our evaluations demonstrate that this method outperforms the current state-of-the-art in equivalence checking, particularly in detecting non-equivalence. Notably, the method scales efficiently with both circuit width and depth when comparing quantum circuits that are closely related, such as when one circuit is derived from the other through compilation with minor errors. While equivalence checking remains a QMA-complete problem, making it intractable for arbitrary circuits, our approach shows significant potential for practical use cases, where we see polynomially-scaling equivalence checking for circuits of realistic size and complexity.

This paper is organized as follows. In Sec. \ref{sec:Preliminaries}, we introduce the essential tensor network building blocks, including basic tensor operations and matrix product operators (MPOs), aimed at readers who may be unfamiliar with tensor networks but are interested in equivalence checking. In Sec. \ref{sec:EquivalenceChecking}, we define the equivalence and identity checking problems, outlining our motivation for addressing them. We then present our algorithm for MPO-based equivalence checking in Sec. \ref{sec:GeneralIdea}, first explaining the approach at a high level. In Sec. \ref{sec:GateApplication}, we detail the tensorization of quantum circuits, the application of individual gates to the intermediary MPO, and the handling of long-range gates. In Sec. \ref{sec:ApplicationStrategy}, we provide the complete implementation strategy, which involves a spatial sweep, utilizing temporal zones of the quantum circuits, as well as the evaluation of equivalence from the resulting MPO. In Sec. \ref{sec:Results}, we introduce the experimental setup and assess the impact of the SVD threshold on our method, as well as evaluate different entanglement patterns to study scaling for both equivalent and non-equivalent circuits against other available techniques. Finally, we discuss the broader implications of our method, propose future research directions, and conclude the work in Sec. \ref{sec:Discussion}.

\section{Preliminaries} \label{sec:Preliminaries}
\begin{figure}[ht!]
    \centering

    \includegraphics[width=0.75\linewidth]{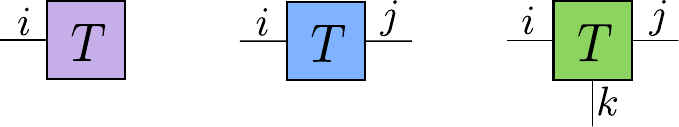}\\
    \textbf{(a)} Example tensor network notation for rank-1, rank-2, rank-3 tensors
    
    \bigskip

    \includegraphics[width=0.5\linewidth]{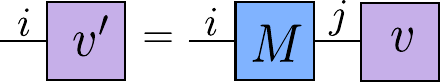}\\
    \textbf{(b)} Contraction of a rank-2 tensor $M_{ij}$ with a rank-1 tensor $v_j$
    (Matrix-vector multiplication, Eq.~\eqref{eq:TensorContraction})
    
    \bigskip

    \includegraphics[width=0.65\linewidth]{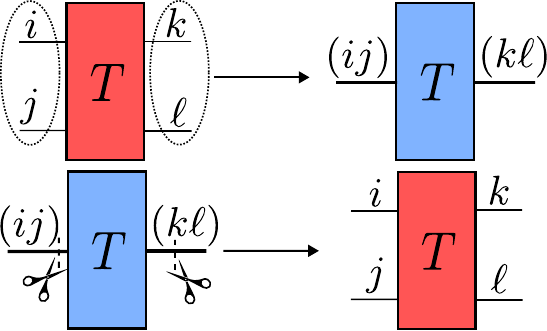}\\
    \textbf{(c)} Reshape of a rank-4 tensor into a rank-2 tensor and vice versa
    (Eq.~\eqref{eq:TensorReshape})
    
    \bigskip

    \includegraphics[width=0.75\linewidth]{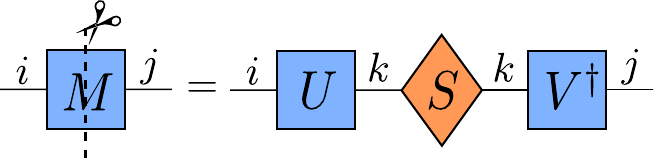}\\
    \textbf{(d)} SVD of an arbitrary matrix $M$ (Eq.~\eqref{eq:TensorSVD});
    the diamond denotes the $S$ matrix

    \bigskip

    \includegraphics[width=0.2\linewidth]{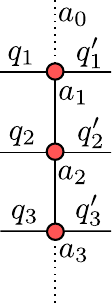}\\
    \textbf{(e)} A 3-site MPO with labeled indices, including $\chi_i=1$ dotted lines
    (Eq.~\eqref{eq:MPOBasics})

    \caption{Basic tensor operations and the MPO structure.  Each panel (a)--(e)
    illustrates one fundamental concept or structure.}
    \label{fig:basic_ops}
\end{figure}

In the past decade, tensor networks have emerged as state-of-the-art computational methods for solving problems in quantum many-body physics, from fundamental theoretical challenges to the simulation of quantum circuits. This section aims to provide readers without a tensor network background with a foundation to understand the methods employed in this work. While this introduction is not exhaustive, it serves as a starting point for comprehending the techniques used in our equivalence checking method. For more in-depth treatments of tensor networks, we refer the reader to several resources \cite{vidal_efficient_2003, schollwoeck_density-matrix_2011, orus_practical_2014, bridgeman_hand-waving_2017, paeckel_time-evolution_2019}. Additionally, a visualization of the operations used throughout this work are found in \autoref{fig:basic_ops}.

Tensors are generalizations of vectors and matrices from linear algebra to multilinear algebra. Computationally, tensors can be viewed as multi-dimensional arrays, with the \textbf{rank} of a tensor referring to the number of dimensions. Scalars, vectors, and matrices are simply rank-0, rank-1, and rank-2 tensors, respectively, while higher-dimensional arrays are referred to as rank-$k$ tensors, where $k \in \mathbb{N}$. Mathematically, a rank-$k$ tensor is indexed by $k$ dimensions and can be visualized as a diagram with $k$ legs extending from a central node. The symbol used to represent a tensor is arbitrary, but multiple symbols may be employed to highlight different roles in a tensor network.

Matrix multiplication can be generalized to \textbf{tensor contraction}, which involves summing over a shared index between two tensors. For example, the matrix-vector multiplication $v' = Mv$ for a vector $v \in \mathbb{C}^b$ and matrix $M \in \mathbb{C}^{a \times b}$ is written as
\begin{equation} 
    v_i' = \sum_{j=1}^n M_{ij} v_j,
    \label{eq:TensorContraction}
\end{equation}
where $v' \in \mathbb{C}^a$, and the contraction is performed over the shared index $j$.

Just as vectors can be reshaped into matrices or vice versa, a \textbf{reshape} operation allows us to transform a rank-$k$ tensor into a rank-$k'$ tensor by combining or splitting dimensions. This operation is essential in tensor networks and is often denoted by grouping indices, such as reshaping a rank-4 tensor $T \in \mathbb{C}^{2 \times 2 \times 2 \times 2}$ into a rank-2 tensor $T \in \mathbb{C}^{4 \times 4}$, written as
\begin{equation}
    T_{ijkl} \rightarrow T_{(ij)(kl)}.
    \label{eq:TensorReshape}
\end{equation}
Reshaping is crucial for tensor network operations, such as decomposition methods, many of which are defined only for matrices. To apply these methods to higher-rank tensors, we first reshape them into matrices, perform the decomposition, and then reshape the results back into the original form (or a desired alternative).

A key decomposition method is the \textbf{singular value decomposition (SVD)}, which decomposes any matrix $M \in \mathbb{C}^{a \times b}$ as
\begin{equation}
    M_{ij} = \sum_{k=1}^\chi U_{ik} S_{kk} V_{kj}^\dagger,
    \label{eq:TensorSVD}
\end{equation}
where $U \in \mathbb{C}^{a \times \chi}$ and $V^\dagger \in \mathbb{C}^{\chi \times b}$ are the left- and right-eigenvector matrices, and $S \in \mathbb{R}^{\chi \times \chi}$ is a diagonal matrix of singular values, with $\chi \leq \min(a, b)$. This process is used extensively in tensor network methods.
The total number of singular values, $\chi$, is referred to as the \textbf{bond dimension}, which represents the amount of information retained between the two parts of the tensor. Truncating $\chi$ based on a threshold $s_{\text{max}}$ yields an approximate decomposition. This method is frequently used on higher-rank tensors, which are reshaped into matrices, decomposed using the SVD, and then contracted or reshaped back into tensors.

The SVD allows for the decomposition of a matrix representing a quantum operator into a tensor train representation known as a \textbf{matrix product operator (MPO)} \cite{pirvu_matrix_2010, hubig_generic_2017, parker_local_2020}. A global operation acting on $n$ qubits, $g \in \mathbb{C}^{2^n \times 2^n}$, can be expressed as an MPO composed of $n$ rank-4 tensors:
\begin{equation}
    W_{q_1, \dots, q_n}^{q'_1, \dots, q'_n} = \sum_{a_0, \dots, a_{n}=1}^{\chi_i} \prod_{i=1}^n W^{a_{i-1}, a_i}_{q_i, q'_i},
    \label{eq:MPOBasics}
\end{equation}
where the physical dimensions $q_i, q'_i$ correspond to local transformations at the $i^{\text{th}}$ qubit, and $a_{i-1}, a_i$ represent the left- and right-bond dimensions. The bond dimensions $\text{dim}(a_i) = \chi_i$ encode the operator’s entanglement capacity.

The values of $\chi_i$ reflect the operator's ability to generate or destroy quantum entanglement \cite{zanardi_entangling_2000, jonay_coarse-grained_2018}. Operators with low bond dimensions $\chi_i$ are less entangling and thus easier to store and manipulate. Efficient storage and computation are possible when $\chi_i$ is small across the network, allowing us to represent and manipulate complex quantum operators without an exponential growth in resources.

\section{Equivalence Checking} \label{sec:EquivalenceChecking}
In the context of quantum computing, equivalence checking involves proving that two quantum circuits, $G$ and $G'$, represent an equivalent unitary operation or demonstrating their non-equivalence via a counterexample. This process is comparable to the verification of classical logic circuits, with one circuit serving as the specification and the other representing the device being verified. Within quantum computing, this approach targets a wide range of use cases, from researchers building circuits by hand who want to ensure they have made no mistakes up to the integration of equivalence checking into quantum computing software stacks which may have multiple levels of compilation and optimization.

\subsection{Considered problem} \label{sec:ConsideredProblem}
Equivalence checking of quantum circuits can be performed using various methods, ranging from simple to complex. To illustrate this, we first define the unitary operators representing two quantum algorithms, $U$ and $U'$, both acting on $n$ qubits. The equivalence checking problem asks whether:
\begin{equation*}
    U = e^{i \theta} U',
\end{equation*}
where $\theta \in (-\pi, \pi]$ denotes a global phase. Unitary operators possess a crucial property that we can exploit:
\begin{equation}
    U U^\dagger = U^\dagger U = I_n,
    \label{eq:UnitaryProperty}
\end{equation}
where $I_n$ is the $n$-qubit identity operator, defined as:
\begin{equation}
    I_n = \underbrace{I \otimes \dots \otimes I}_{n},
    \label{eq:Identity}
\end{equation}
and $I$ is the local identity operator. One of the simplest ways to check whether two unitaries $U$ and $U'$ are equivalent is to determine if they satisfy Eq. \eqref{eq:UnitaryProperty} when substituted for each other:
\begin{equation}
    U U'^\dagger = U' U^\dagger = e^{i \theta} I_n.
    \label{eq:UnitaryProperty2}
\end{equation}

However, in quantum computing, the entire unitary operator $U$ (and similarly $U'$) is typically not directly accessible for comparison. Instead, the unitary must be reconstructed from the individual quantum gates:
\begin{equation*}
    G[U] = g_0 \dots g_{|G|-1},
\end{equation*}
where $G[U]$ represents a quantum circuit encoding $U$, composed of $|G|$ gates $g_i$ for $0 \leq i < |G|$. For brevity, we will drop the bracket notation of $G[U]$.

Given the reversibility of quantum algorithms (and the properties of unitary operators in Eq. \eqref{eq:UnitaryProperty} and Eq. \eqref{eq:UnitaryProperty2}), the equivalence of two quantum circuits, $G$ and $G'$, can be determined using the following equivalence checking condition:
\begin{equation}
    G = G' \Longleftrightarrow G G'^\dagger = e^{i \theta} I_n,
\end{equation}
where $G'^\dagger$ denotes the conjugation and reversal of all gates in $G'$. Thus,
\begin{equation}
    G G'^\dagger = g_0 \dots g_{|G|-1} (g'_{|G'|-1})^\dagger \dots (g'_0)^\dagger,
    \label{eq:Gates}
\end{equation}
which leads to two QMA-complete problems \cite{janzing_identity_2003}: the \textbf{identity check}
\begin{equation}
    G G'^\dagger \stackrel{?}{=} I_n,
    \label{eq:IdentityCheck}
\end{equation}
and the \textbf{equivalence check}
\begin{equation}
    G \stackrel{?}{=} G'.
    \label{eq:EquivalenceCheck}
\end{equation}

To solve this computationally, each gate $g_i$ must be stored and manipulated to construct $G G'^\dagger$. In its simplest form, this can be done using matrix-matrix multiplication, where each qubit-specific operation is extended using the tensor product to create global matrices $g_i \in \mathbb{C}^{2^n \times 2^n}$. However, Eq. \eqref{eq:Gates} suffers from exponential growth in both the size of the matrices and the computational cost of performing the equivalence check. Therefore, it becomes necessary to develop methods that mitigate this exponential scaling, at least in practical cases, by encoding gate operations in more compact data structures rather than relying solely on matrices.

\subsection{Related work} \label{sec:RelatedWork}
Several data structures have been developed to mitigate the exponential scaling problems in quantum circuit equivalence checking, including the ZX-Calculus, decision diagrams (DDs), and tensor decision diagrams (TDDs). Each of these methods, however, has distinct limitations that we aim to address.

The \textbf{ZX-Calculus}~\cite{vandewetering2020zxcalculus,cowtan_generic_2020,duncan_graph-theoretic_2020, kissinger_reducing_2020} is a graphical notation for quantum circuits, equipped with a powerful set of rewrite rules that enable diagrammatic reasoning about quantum systems. Primarily used for circuit compilation and optimization, ZX-Calculus has also been applied to equivalence checking~\cite{Peham_2022}. In this approach, one of the circuits is inverted and combined with the other, forming $G G'^\dagger$, as described by Eq. \eqref{eq:Gates}. The rewrite rules are then used to simplify the combined ZX diagram. If the diagram reduces to bare wires, the circuits are deemed equivalent. This method is efficient for large circuits but has a major drawback: the ZX-Calculus, as proposed in \cite{Peham_2022}, is incomplete. It cannot reliably conclude non-equivalence if the diagram does not fully reduce to the identity~\cite{backens_zx-calculus_2014, de_witt_zx-calculus_2014}.

Another approach is the use of \textbf{decision diagrams (DDs)}~\cite{zulehner_advanced_2019, wille_tools_2022, wille_decision_2023}, which are inspired by binary decision diagrams (BDDs) from classical computing~\cite{minato_zero-suppressed_1993}. DDs represent quantum circuits as Directed Acyclic Graphs (DAGs) with complex-valued edge weights by recursively splitting matrices into sub-parts and exploiting redundancies. Applied to equivalence checking~\cite{burgholzer_advanced_2021}, an intermediary identity operator $I_n$, represented by a DD, is placed between two circuits $G$ and $G'^\dagger$, as described in Eq. \eqref{eq:Gates}. Gates from each circuit are alternately applied to the intermediary DD, which remains compact if the gates effectively cancel each other out, leaving a structure equivalent to $G G'^\dagger$, which can be compared to $I_n$ to determine equvialence or non-equivalence. However, DDs can become exponentially large in the worst case, particularly when circuits are non-equivalent, as the resulting DD may not reduce to the compact identity diagram~\cite{sander_stripping_2024}.

\textbf{Tensor decision diagrams (TDDs)}~\cite{hong_approximate_2021, hong_equivalence_2021, hong_equivalence_2024} offer another approach, combining ideas from tensor networks with DD-like data structures. Each gate is reduced to a local tensor operation on the qubits it interacts with, and the result is stored in a DD format to exploit redundancies. This method contracts gates from $G$ and $G'^\dagger$, followed by tracing the TDD to perform the identity check. However, it suffers from the same limitations as DDs, such as the potential for exponential growth in non-equivalent cases. Additionally, the TDD approach does not fully utilize many of the advanced techniques developed for tensor network simulation, focusing instead on exploiting redundancies within the TDD format.

Alternatively, all operations in the circuits $G$ and $G'^\dagger$ could be directly represented as tensors, and their equivalence could be checked using standard tensor contraction algorithms~\cite{gray_hyper-optimized_2021}. In this approach, the full network representing $G G'^\dagger$ would be contracted. However, this method faces scalability challenges, as the intermediate tensors generated during contraction, as well as the final tensor, scale exponentially with the number of qubits, similar to the issues faced by the TDD approach.

\section{MPO-based Equivalence Checking} \label{sec:GeneralIdea}
To address the limitations of the methods described in previous works, we propose an equivalence checking method based on tensorized quantum circuits \cite{huang2020classical, gray_hyper-optimized_2021, pan_simulation_2022} and an intermediary matrix product operator (MPO) \cite{pirvu_matrix_2010, hubig_generic_2017, parker_local_2020} that follows an alternating strategy: $G \rightarrow I_n \leftarrow G'^\dagger$. Unlike decision diagrams (DDs), where the entire system’s qubits are interdependent, the local tensors of an MPO ensure that operations only affect local regions, avoiding global growth in complexity. Additionally, this MPO-based method can conclusively prove non-equivalence, unlike the incomplete ZX-based approach from \cite{Peham_2022}.

Each quantum gate can be reduced to local operations that act on specific qubits. For $k$-qubit gates (kQGs), this reduction is done by reshaping the matrix representation into a tensor form. While MPO methods are generally designed for nearest-neighbor qubits, long-range gates (which are essential for many quantum algorithms) can be similarly represented by decomposing them and then extending the tensors into an MPO form using identity tensors. This approach decouples the gate representation from the total qubit count, resulting in a compact tensor representation.

The proposed MPO-based equivalence checking algorithm leverages this tensor representation of quantum gates. To handle the scalability challenges of directly contracting tensors from $G G'^\dagger$, we introduce an intermediary MPO representing the identity operation, akin to the DD-based strategy \cite{burgholzer_advanced_2021} mentioned in Sec. \ref{sec:RelatedWork}, but with a novel data structure and application strategy. By inserting the intermediary MPO between the circuits, we can represent the operation as:
\begin{equation*}
    G I_n G'^\dagger = g_0 \dots g_{|G|-1} I_n (g'_{|G'|-1})^\dagger \dots (g_0)^\dagger,
\end{equation*}
and contract gates from each circuit into the MPO according to the following application strategy:
\begin{equation*}
    G \rightarrow I_n \leftarrow G'^\dagger.
    \label{eq:GateApplication}
\end{equation*}
Since the $n$-qubit identity matrix $I_n$ is formed as a repeated tensor product of local identity matrices (Eq. \eqref{eq:Identity}), its MPO representation is maximally compact with bond dimensions $\chi_i = 1 \ \forall i$. The tensorized gates are applied to the local tensors of the MPO by contracting their physical indices $q_i, q'_i$, corresponding to the inputs for gates from $G$ and $G'^\dagger$, respectively.

The gates are applied sequentially in a two-qubit spatial sweep across the tensors of the intermediary MPO. For each qubit pair (spatial zone), we apply all gates from both circuits that act within that spatial zone (temporal zone). This strategy minimizes the number of singular value decompositions (SVDs) needed and helps keep the bond dimensions as low as possible, especially in cases where the circuits are equivalent or nearly equivalent.

After all gates have been applied and the intermediary MPO $W$ represents $GG'^\dagger$, we can check the equivalence of the circuits by calculating the Frobenius inner product with the identity:
\[
\langle I_N, GG'^\dagger \rangle = \langle I_N, W \rangle = \text{Tr}[W].
\]
This enables efficient evaluation of the equivalence condition from Eq. \eqref{eq:EquivalenceCheck} and results in a complete MPO-based equivalence checking algorithm.

\begin{figure*}[t]
    \centering
    
    \includegraphics[width=0.3\textwidth]{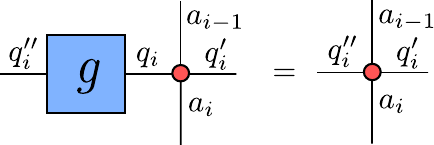}\\
    \textbf{(a)} Application of a 1QG to an MPO
    
    \bigskip

    \includegraphics[width=0.8\textwidth]{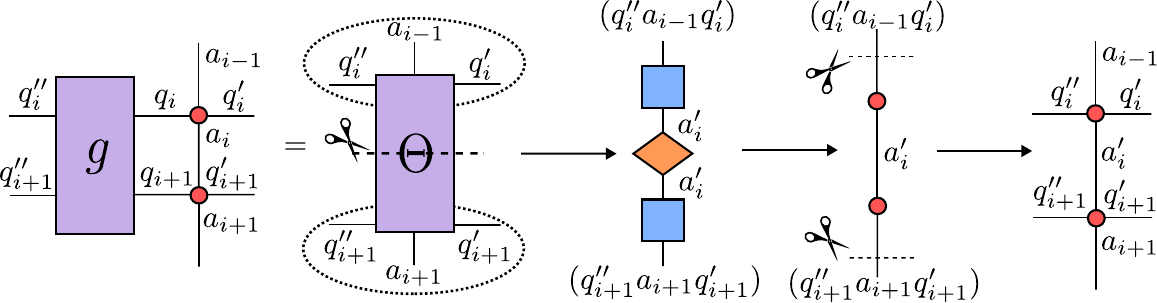}\\
    \textbf{(b)} Application of a 2QG to an MPO

    \bigskip

    \includegraphics[width=0.9\textwidth]{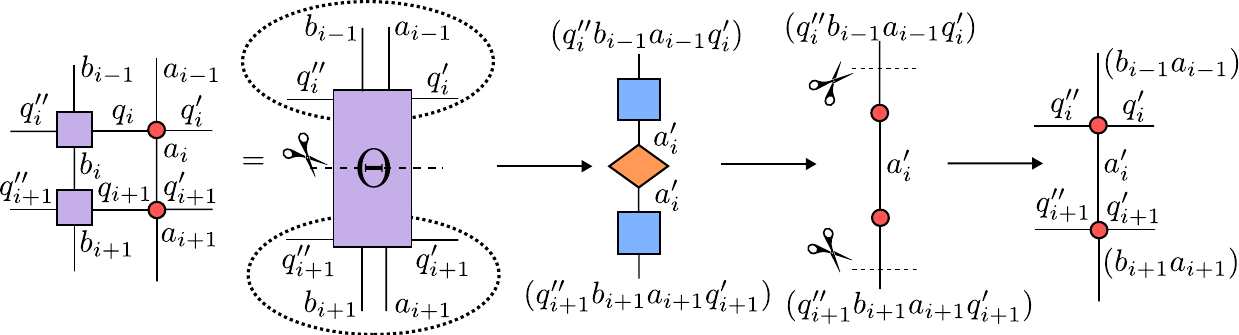}\\
    \textbf{(c)} Application of two sites of a long-range gate to an MPO

    \caption{Visualization of MPO operations required for the equivalence checking algorithm.}
    \label{fig:TensorOperations}
\end{figure*}

The following section details the implementation for readers less familiar with tensor network methods, particularly how quantum circuit gates are initialized and applied to the MPO structure. These operations are explained through equations using tensor notation and are visualized in \autoref{fig:TensorOperations}. Readers already familiar with this topic may proceed to Sec. \ref{sec:ApplicationStrategy}, where we discuss the application strategy that underpins the computational efficiency of the MPO-based method.

\section{Gate Application} \label{sec:GateApplication}
To perform MPO-based equivalence checking, quantum gates must be applied to the intermediary MPO. This involves tensorizing the gates and determining how and where they are applied within the local site tensors of the MPO. For kQGs with $k \leq 2$ acting on neighboring qubits, this process is straightforward. However, for kQGs with $k > 2$ and long-range gates, extra care is needed, as MPOs are optimized for nearest-neighbor interactions. This section describes how quantum gates are represented as tensors and how individual gates are applied to the intermediary MPO.

\subsection{Nearest-neighbor gates}
Any kQG can be represented by a rank-$2k$ tensor, obtained by reshaping the matrix form into local dimensions corresponding to each site:
\begin{equation*}
    g \in \mathbb{C}^{2^k \times 2^k} \rightarrow g \in \mathbb{C}^{\overbrace{2 \times \dots \times 2}^{2k}}.
\end{equation*}
For a 2QG, this reshaping creates a rank-4 tensor: $g \in \mathbb{C}^{4 \times 4} \rightarrow \mathbb{C}^{2 \times 2 \times 2 \times 2}$. Each dimension corresponds to the local input and output leg of the tensor for each qubit.

When a gate is applied to neighboring qubits, its tensor can be applied directly to the local tensors of the intermediary MPO. For a single-qubit gate (1QG) $g \in \mathbb{C}^{2 \times 2}$ acting on qubit $i$ from algorithm $G$, it is contracted into the MPO $W$ at site $i$ along the corresponding index $q_i$:
\begin{equation}
    W^{a_{i-1}, a_i}_{q''_i, q'_i} = \sum_{q_i=0}^1 g_{q''_i, q_i} W^{a_{i-1}, a_i}_{q_i, q'_i},
\end{equation}
where $\text{dim}(q_i) = \text{dim}(q'_i) =\text{dim}(q''_i)$. Similarly, a 1QG from $G'^\dagger$ acts on index $q'_i$. Since these gates are local, they do not generate entanglement and do not increase the bond dimensions of the MPO tensors. This process is illustrated in \autoref{fig:TensorOperations}\textbf{(a)}.

For kQGs with $k \geq 2$, the application requires contraction of all tensors they act upon, including bond dimensions, forming a higher-rank tensor. For example, a 2QG, represented as a rank-4 tensor $g \in \mathbb{C}^{2 \times 2 \times 2 \times 2}$ acting on neighboring qubits $i$ and $i+1$, results in a rank-6 tensor:
\begin{equation}
    \Theta_{q''_i, q''_{i+1}, q'_i, q'_{i+1}}^{a_{i-1}, a_{i+1}} = \sum_{\substack{q_i, q_{i+1} \\ a_i}} g_{q''_i, q''_{i+1}}^{q_i, q_{i+1}} W^{a_{i-1}, a_i}_{q_i, q'_i} W^{a_i, a_{i+1}}_{q_{i+1}, q'_{i+1}}.
    \label{eq:TwoQubitApplication}
\end{equation}
This tensor is reshaped into a matrix, grouping dimensions across the qubits:
\begin{equation*}
    \Theta_{q''_i, q''_{i+1}, q'_i, q'_{i+1}}^{a_{i-1}, a_{i+1}} \rightarrow \Theta^{(a_{i-1} q''_i q'_i)}_{(a_{i+1}, q''_{i+1}, q'_{i+1})} := \Theta_{\text{mat}},
    \label{eq:2QubitSVD}
\end{equation*}
and then an SVD is applied:
\begin{equation}
    \Theta_{\text{mat}} = U (S V^\dagger) =: W^{(a_{i-1} a'_i q''_i q'_i)} W^{(a'_i a_{i+1} q''_{i+1} q'_{i+1})},
\end{equation}
resulting in a new bond dimension $a_i \rightarrow a'_i$. The singular values in $S$ can be truncated if necessary before multiplying it into one of the neighboring matrices. Finally, the resulting matrices are reshaped into rank-4 tensors for the intermediary MPO:
\begin{equation*}
    W^{a_{i-1}, a'_i}_{q''_i, q'_i} W^{a'_i, a_{i+1}}_{q''_{i+1}, q'_{i+1}}.
\end{equation*}
This process is visualized in \autoref{fig:TensorOperations}\textbf{(b)}.

For kQGs with $k > 2$, this process results in larger intermediate tensors $\Theta$, which require repeated applications of the SVD, and more opportunites for the bond dimension to grow. To avoid this, we decompose kQGs with $k > 2$ into MPOs and treat them as long-range gates to prevent the creation of large $\Theta$ tensors.

\subsection{Long-range gates}
Any kQG can be decomposed into a $k$-site MPO using repeated SVDs. Each site of this MPO represents a local interaction, while the bond dimension allows information to flow between tensors. For example, a 2QG can be decomposed as:
\begin{equation*}
    g_{mn} = \sum_{b_j=1}^\chi A_{m, b_j} B_{b_j, n},
\end{equation*}
where $A \in \mathbb{C}^{4 \times \chi}$ and $B \in \mathbb{C}^{\chi \times 4}$. These tensors can be reshaped into local tensors of an MPO: $A \rightarrow \mathbb{C}^{2 \times 2 \times 1 \times \chi}$ and $B \rightarrow \mathbb{C}^{2 \times 2 \times \chi \times 1}$.

This leads to a nearest-neighbor 2-site MPO representation of the gate:
\begin{equation*}
    g_{q_j, q_{j+1}}^{q'_j, q'_{j+1}} = \sum_{b_j=1}^\chi A^{b_{j-1}, b_j}_{q_j, q'_j} B^{b_j, b_{j+1}}_{q_{j+1}, q'_{j+1}},
\end{equation*}
where $b_{j-1} = b_{j+1} = 1$, and the bond dimension $\chi$ is determined by the properties of the decomposition. This can be extended for long-range gates acting on non-neighboring qubits $j$ and $\overline{j}$ by inserting identity tensors at non-interacting sites:
\begin{equation*}
    \begin{split}
        g^{q'_j, \dots, q'_{\overline{j}}}_{q_j, \dots, q_{\overline{j}}} &= \sum_{b_j, \dots, b_{\overline{j}}=1}^{\chi} A^{b_{j-1}, b_j}_{q_j, q'_j} \ \biggl[ \prod_{i=j+1}^{\overline{j}-1} I^{b_{i-1}, b_i}_{q_i, q'_i} \biggr] B^{b_{\overline{j}-1}, b_{\overline{j}}}_{q_{\overline{j}}, q'_{\overline{j}}} \\
        &=: \sum_{b_j, \dots, b_{\overline{j}}=1}^{\chi}  \ \prod_{i=j}^{\overline{j}} g^{b_{i-1}, b_i}_{q_i, q'_i}.
    \end{split}
    \label{eq:LongRangeGate}
\end{equation*}

Long-range gates are applied to the intermediary MPO by iteratively contracting the gate MPO's tensors in a sweep. For a long-range gate $g$ acting on sites $j$ to $\overline{j}$, we start at qubits $(j, j+1)$ and create a two-site tensor $\Theta$ as in Eq. \eqref{eq:TwoQubitApplication}:
\begin{equation*}
    \begin{split}
        \Theta_{q''_j, q''_{j+1}, q'_j, q'_{j+1}}^{a_{j-1}, b_{j-1}, a_{j+1}, b_{j+1}} =  \sum_{\substack{q_j, q_{j+1} \\ a_j, a'_j}}
        & g^{b_{j-1}, b_j}_{q''_j, q_j} g^{b_{j}, b_{j+1}}_{q''_{j+1}, q_{j+1}} \\
        & W^{a_{j-1}, a_j}_{q_j, q'_j} W^{a_{j}, a_{j+1}}_{q_{j+1}, q'_{j+1}}. 
    \end{split}
\end{equation*}
After constructing $\Theta$, it is reshaped and decomposed using the SVD, similar to the nearest-neighbor case. This process is repeated for all sites until the end of the gate at $\overline{j}$ is reached, at which point the long-range gate MPO has been fully applied to the intermediary MPO. This is visualized in Fig. \autoref{fig:TensorOperations}\textbf{(c)}.

Applying gates with $k > 1$ requires SVD decompositions to return the intermediate tensor $\Theta$ to the original MPO form. The SVD is the most computationally expensive step in the equivalence checking algorithm and can cause bond dimension growth, potentially leading to slowdowns. To mitigate this, we aim to extend the individual gate application to a multi-gate application strategy, allowing us to apply as many operations to $\Theta$ as possible before performing the decomposition.

\section{Application Strategy} \label{sec:ApplicationStrategy}
\begin{figure*}[t]
    \centering

    \includegraphics[width=\linewidth]{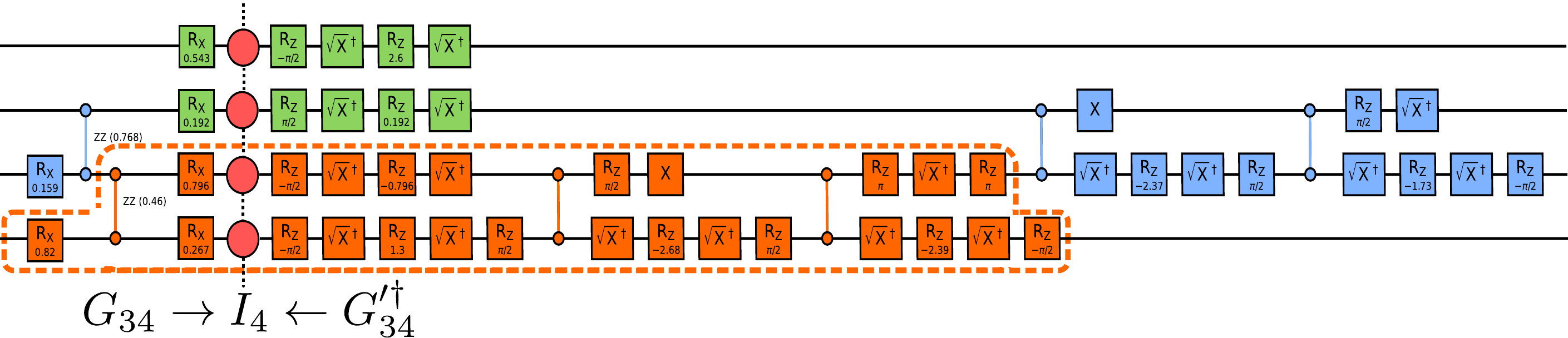}\\
    \textbf{(a)} This figure shows an example of two equivalent linear 
    entanglement algorithms with an intermediary MPO. Each color 
    represents the zones during a sweep with the order 
    $G_{12} \rightarrow I_4 \leftarrow G'^\dagger_{12}$ (green), 
    $G_{34} \rightarrow I_4 \leftarrow G'^\dagger_{34}$ (orange, 
    encircled as an example), then 
    $G_{23} \rightarrow I_4 \leftarrow G'^\dagger_{23}$ (blue).

    \bigskip

    \includegraphics[width=\linewidth]{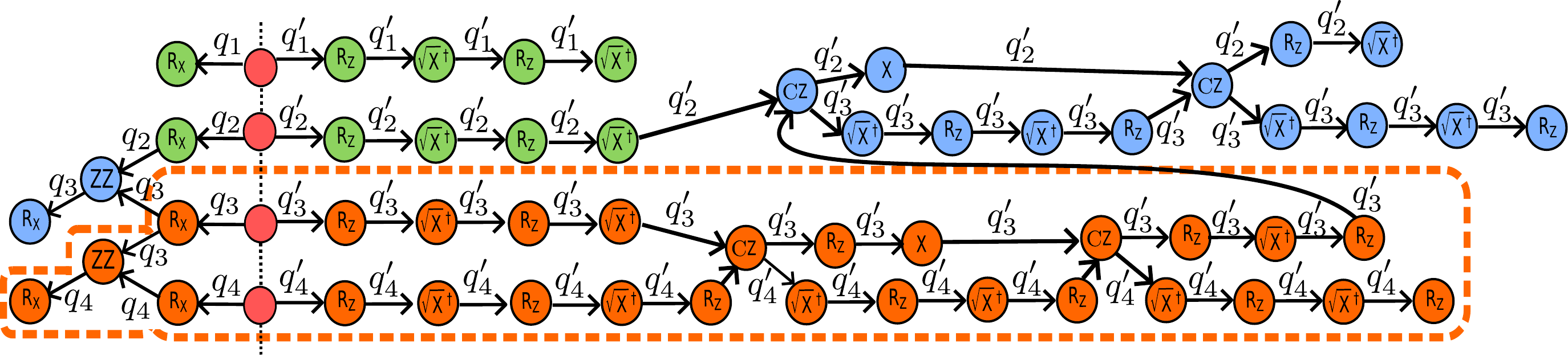}\\
    \textbf{(b)} This shows the equivalent DAG representation of the 
    above circuit. The transition edges represent the qubit it corresponds 
    to, allowing easy identification of zones and the order of tensors to apply.

    \caption{Visualization of the MPO-based equivalence checking algorithm in 
    both quantum circuit (a) and DAG (b) form.}
    \label{fig:FullAlgorithmCausalCones}
\end{figure*}

In previous works, the strategy for applying gates between circuits has been crucial in maintaining the compactness of the intermediary data structure, whether by applying gates one-to-one, proportionally based on the number of gates, or using more advanced strategies \cite{burgholzer_advanced_2021}. While such strategies can benefit from the spatial locality of quantum circuits, they often do not fully exploit the circuit's temporal structure. Therefore, instead of reusing these strategies from other data structures, we propose a new application strategy that fully leverages the advantages offered by tensor networks.

\subsection{Spatial and temporal zones}
Our approach is based on a two-site spatial sweep across pairs of qubits, which is common in tensor network methods such as TEBD \cite{vidal_efficient_2003} and DMRG \cite{schollwoeck_density-matrix_2011}. In the proposed MPO-based equivalence checking, we use this sweep to systematically update the tensors of the intermediary MPO by iterating over pairs of qubits, breaking down the quantum circuit into smaller, more manageable chunks—similar to a double-sided, MPO-based TEBD algorithm.

However, unlike previous proportional gate application strategies, we simultaneously apply gates from both circuits within what we call the "temporal zone" of the qubits. This zone includes all gates that act exclusively on the selected pair of qubits and is conceptually similar to the causal (or light) cones used in expectation value calculations \cite{frias-perez_light_2022, anand_information_2022, shehab_noise_2019, ran_tensor_2020}. The temporal zone application digs as deeply into the circuits as possible, applying as many gates as available at each step. Instead of requiring an SVD after each multi-qubit gate, this strategy requires an SVD only after applying the full temporal zone, which minimizes the number of SVDs required and helps prevent bond dimension growth, particularly when the circuits are equivalent or nearly equivalent.

\subsection{Combined sweep}
In this method, we perform a two-site spatial sweep across the intermediary MPO \(W\), alternating between even-indexed and odd-indexed qubits. At each spatial zone \((i, i+1)\), we apply tensorized gates from both circuits, \(G\) and \(G'^\dagger\), in the order defined by the directed acyclic graph (DAG) representation of each circuit.

The DAG representation provides a structured way to identify temporal zones, where we traverse the edges corresponding to qubits \((i, i+1)\) and sequentially apply gates until encountering a node connected to another qubit outside the current spatial zone. This traversal marks the end of the temporal zone for \((i, i+1)\). Nodes corresponding to applied gates are removed from the DAG, and the process is repeated for subsequent spatial zones. As shown in \autoref{fig:FullAlgorithmCausalCones}, the DAG representation helps determine the temporal zones and their associated gate sequences.

Differences in the DAG structures of \(G\) and \(G'\) can introduce mismatches in the temporal zones, particularly when long-range gates or differences in qubit connectivity are present. For example, while the transpilation process for \(G'\) may assume all-to-all qubit connectivity, \(G\) may retain a specific structure such as a linear nearest-neighbor structure. This mismatch results in temporal zones in \(G'\) that may not perfectly align with those in \(G\). In such cases, long-range gates in either circuit require recursive treatment.

More precisely, to ensure efficient operator cancellations even in cases of long-range gates, the method introduces an MPO-MPO contraction step when a long-range gate or \(k\)-qubit gate (\(k > 2\)) is encountered. Here, the overall even/odd sweep is interrupted to handle the long-range interaction by using a subsweep along the sites of the gate MPO. As each set of neighboring sites in the gate MPO are contracted with the intermediary MPO, the temporal zone is applied from deeper into each circuit. This approach minimizes the computational cost associated with handling mismatched DAG structures while preserving the operator cancellations enabled by local temporal zones. By addressing potential mismatches in the DAG structures of \(G\) and \(G'\) through recursive subzone processing, this method ensures robustness and efficiency even in the presence of long-range gates or varying circuit connectivity patterns.

Once all gates are applied for a given spatial zone, the resulting tensor \(\Theta = G_{i, i+1} W_i W_{i+1} G'^\dagger_{i, i+1}\) is decomposed using the SVD, reshaped back into two rank-4 tensors, and used to update the intermediary MPO. The even/odd sweep continues across the MPO until all gates have been applied and all nodes removed from the DAG.

\subsection{Evaluating $G G'^\dagger$} \label{sec:Evaluation}
\begin{figure}
    \centering
    \includegraphics[width=0.25\linewidth]{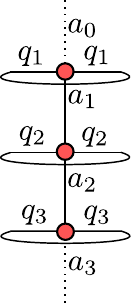}
    \caption{Tensor trace of an MPO (Eq. \eqref{eq:TensorTrace}) used to perform the identity check in Eq. \eqref{eq:TraceBound}}
    \label{fig:TensorTrace}
\end{figure}

Once all gates have been contracted into the intermediary MPO, the resulting MPO $W$ now encodes $G G'^\dagger$. The equivalence of $G$ and $G'$ can be evaluated using Eq. \eqref{eq:EquivalenceCheck}. Specifically, we check how close $W$ is to the identity operation $I_n$ by calculating the Frobenius inner product $\langle A, B \rangle_F = \text{Tr}[A^\dagger B]$. The inner product between $W$ and $I_n$ (possibly with a global phase) simplifies to:
\[
    \langle e^{-i \theta} I_n, W \rangle_F = e^{i \theta} \ \text{Tr}[W],
\]
which can be computed by contracting the physical dimensions $q_i$ and $q'_i$:
\begin{equation}
    \text{Tr}[W] = \sum_{a_0, \dots, a_n=1}^{\chi_i} \sum_{q_1, \dots, q_n=0}^{d-1} \prod_{i=1}^n W^{a_{i-1}, a_i}_{q_i, q_i}.
    \label{eq:TensorTrace}
\end{equation}
This process is illustrated in \autoref{fig:TensorTrace}.

Unlike quantum states, operators do not inherently require normalization; thus, the trace of \(W\) may not be 1. However, for an \(n\)-qubit identity operator \(I_n\), we have \(\mathrm{Tr}[I_n] = 2^n\), making \(I_n\) (up to a global phase) the unique unitary achieving that maximum trace. Consequently, we can check equivalence via
\[
    \biggl|\,e^{i\theta}\,\frac{\mathrm{Tr}[W]}{2^n}\biggr| \approx 1 
    \quad\Longleftrightarrow\quad G = G',
\]
which in practice leads to the condition
\begin{equation}
    \biggl|\frac{\mathrm{Tr}[W]}{2^n}\biggr| \;\ge\; 1 - \epsilon 
    \quad\Longleftrightarrow\quad G = G',
    \label{eq:TraceBound}
\end{equation}
where \(\epsilon\) (\(0 \le \epsilon < 1\)) defines a numerical tolerance such that 
\[
  1 - \epsilon \;\le\; \frac{\mathrm{Tr}[W]}{2^n} \;\le\; 1 + \epsilon.
\]
Furthermore, the global phase does not affect \(\tfrac{\mathrm{Tr}[W]}{2^n}\) because 
\[
  \biggl|\,e^{i\theta}\,\frac{\mathrm{Tr}[W]}{2^n}\biggr|
  \;=\;
  |\,e^{i\theta}\,|\,
  \biggl|\frac{\mathrm{Tr}[W]}{2^n}\biggr|
  \;=\;
  \biggl|\frac{\mathrm{Tr}[W]}{2^n}\biggr|.
\]

This identity check can be scaled to any length MPO, enabling the use of an early stopping condition. For $W$ to be equivalent to the $n$-qubit identity $I_n$, each local tensor $W_i$ must be equivalent to the local identity $I$. Therefore, if all gates acting on a specific qubit $i$ have been applied, but the partial trace reveals non-equivalence for that qubit, we can conclude that the circuits are not equivalent and terminate the algorithm early. Depending on the circuit topology, this early stopping condition can significantly speed up the algorithm. This could also be extended to detect specific circuit bugs, although this is left for future work.

An alternative way to verify \(GG'^\dagger\) is to first check whether all bond
dimensions \(\chi_i = 1\) for all $i$, then check if each local matrix is the identity. However, in practice, small numerical
imprecisions often cause slight increases in bond dimensions, especially for
larger systems, even when the result is the identity. A final SVD sweep
across the entire MPO can reduce these extra dimensions, but it is
computationally less efficient than directly calculating
\(\mathrm{Tr}[GG'^\dagger]\). Moreover, \(\chi_i = 1\) could also mean that the
operator factorizes into arbitrary local operators, making this criterion ambiguous. By contrast, computing the trace
immediately provides a clear global check of equivalence that is both more
robust and efficient in most practical settings.

\subsection{Advantages}
This application strategy offers two key benefits. First, it reduces the number of SVDs required by applying multiple gates at once, ensuring that if several 2QGs lie within the temporal zone, only a single SVD is needed. This can lead to significant efficiency gains, depending on the circuit structure. Most importantly, by applying gates from both circuits simultaneously, the bond dimension remains small when the circuits are structurally similar, such as when one is derived from the other via compilation. In such cases, the circuits effectively cancel out one another, reducing the operator entanglement and keeping the MPO bond dimensions low throughout the algorithm, ultimately improving computational efficiency.

\section{Results} \label{sec:Results}
To evaluate the performance and scalability of the proposed MPO-based equivalence checking method, we implemented a Python prototype using numpy \cite{harris2020array} for tensor representations and opt-einsum \cite{smith_opt_einsum_2018} for efficient tensor operations. This implementation is part of the \mbox{\emph{Munich Quantum Toolkit (MQT)}} \cite{wille_mqt2024} and is available in the \emph{YAQS} package found at~\cite{YAQS}, as well as integrated as a tool in \emph{QCEC} \cite{burgholzer_advanced_2021}.
The performance of this method is compared against the established DD- and ZX-based methods found in the QCEC package. All tests were executed on an i7-1165G7 (2.80 GHz).

For benchmarking, we generated parameterized two-local quantum circuits using IBM's qiskit \cite{qiskit2024}. Each circuit consists of a block made of a layer of single-qubit rotation gates, $R_x$, followed by an entanglement layer of two-qubit rotation gates $R_{zz}$ (with various entanglement patterns). This is then repeated before a final layer of $R_x$ gates.

The tested circuits are then generated as square \(n \times n\) circuits, meaning that for \(n\) qubits, there are \(n\) repetitions of this rotation structure. Each rotation angle, \(\theta_i\), is randomly (according to a uniform distribution) selected from the range \([- \pi, \pi]\), with each block generated independently such that the structures do not repeat. This is visualized in \autoref{fig:ExperimentalSetup}\textbf{(a)}.

To evaluate the performance of the MPO-based method, we consider three classes of entanglement patterns, each distinguished by how many long-range gates they contain:
\begin{enumerate}
    \item \textbf{Linear entanglement}\\
    Each qubit interacts only with its nearest neighbors. In each layer,
    the \(R_{zz}\) gates follow a staircase pattern, connecting \((q_i, q_{i+1})\) for $i=1, \dots, n$.. 
    Hence, there are no long-range gates. This is shown in \autoref{fig:ExperimentalSetup}\textbf{(b)}

    \item \textbf{Shifted-circular alternating (SCA) entanglement}\\
    This pattern extends the linear circuit by adding a single long-range 
    gate \((q_1, q_{n})\) within each layer. The order of gates in a layer 
    shifts so that the long-range gate alternates its position each time 
    \cite{Sim_2019}. For example, in one layer the entangling gates contains a long-range gate \((q_1, q_n)\) followed by \((q_i, q_{i+1})\) for $i=1, \dots, n$. In the next layer, this long-range gate shifts such that the circuit has a gate \((q_1, q_2)\), then the long-range  \((q_1, q_n)\), followed by \((q_i, q_{i+1})\) for $i=2, \dots, n$. The first repetition block is shown in \autoref{fig:ExperimentalSetup}\textbf{(c)} as an example.

    \item \textbf{Full entanglement}\\
    Each qubit interacts with every other qubit in each layer, resulting in 
    many long-range gates. Specifically, the \(R_{zz}\) gates connect all 
    pairs \((q_i, q_{j})\) for $i,j=1, \dots, n$ where $i \neq j$
    creating both a larger total number of gates and extensive long-range 
    entanglement. This is shown in \autoref{fig:ExperimentalSetup}\textbf{(d)}.
\end{enumerate}
These circuits were chosen to provide a generalized set of quantum circuits, representing a wide-range of applications, suitable for thoroughly testing the proposed equivalence checking method.

The effectiveness of MPO-based methods, and similar MPS-based methods, is heavily influenced by the interaction range. These classes are designed to represent a spectrum of scenarios, ranging from ideal (Linear entanglement) to non-ideal (Full entanglement), with a transitional regime (SCA entanglement) in between.

After generating the circuits \(G\), each circuit is transpiled—optimizing and mapping the gates—to a new circuit \(G'\) with the gate set supported by IBM's Heron architecture:
\[
    \{I, X, \sqrt{X}, R_z(\theta), \text{CZ}\},
\]
with all-to-all qubit connectivity, allowing long-range gates to exist in \(G'\). We then use the circuits \(G\) and \(G'\) to test each equivalence checking method, either by directly comparing the circuits or by introducing deliberate errors into \(G'\). Specifically, we consider three types of errors:

\begin{enumerate}
    \item \textbf{Missing gate error} \\ A number of gates are randomly removed from \(G'\), representing faulty transpiler passes.
    
    \item \textbf{Rotation angle error} \\ All rotation angles in \(G'\) are offset by a fixed amount, simulating rounding errors or uncalibrated control hardware.
    
    \item \textbf{Permutation error} \\ Random SWAP gates are inserted at the beginning of \(G'\), representing mismatched virtual-to-physical qubits or incorrect qubit ordering.
\end{enumerate}
For all tests, each data point represents the average runtime over 10 samples with a trace fidelity tolerance of \(\epsilon = 10^{-13}\). The average runtime \(\overline{T}\) and the standard deviation \(\sigma(T)\) are reported for each test case.

\begin{figure}[ht!]
    \centering
    \begin{minipage}[t]{0.45\textwidth}
        \centering
        \resizebox{\linewidth}{!}{
            \begin{tikzpicture}
                \begin{yquantgroup}
                    \registers {
                        qubit {} q[4];
                    }
                    \circuit{
                        box {$R_{x}(\theta_1)$} (q[0]);
                        box {$R_{x}(\theta_2)$} (q[1]);
                        box {$R_{x}(\theta_3)$} (q[2]);
                        box {$R_{x}(\theta_4)$} (q[3]);

                        box {Entanglement layer} (q[0], q[1], q[2], q[3]);

                        box {$R_{x}(\theta_5)$} (q[0]);
                        box {$R_{x}(\theta_6)$} (q[1]);
                        box {$R_{x}(\theta_7)$} (q[2]);
                        box {$R_{x}(\theta_8)$} (q[3]);
                    }
                \end{yquantgroup}
            \end{tikzpicture}
        }\\[6pt]
        \textbf{(a)} This visualizes the general circuit structure used in all 
        experimental setups. This structure is repeated \(n\) times, each time 
        with different random rotation angles, using the entanglement layers below.
    \end{minipage}

    \vspace{0.8cm}

    \begin{minipage}[t]{0.45\textwidth}
        \centering
        \resizebox{\linewidth}{!}{
            \begin{tikzpicture}
                \begin{yquantgroup}
                    \registers {
                        qubit {} q[4];
                    }
                    \circuit{
                        box {$R_{zz}(\theta_1)$} (q[0], q[1]);
                        box {$R_{zz}(\theta_2)$} (q[1], q[2]);
                        box {$R_{zz}(\theta_3)$} (q[2], q[3]);
                    }
                \end{yquantgroup}
            \end{tikzpicture}
        }\\[6pt]
        \textbf{(b)} Linear entanglement layer
    \end{minipage}

    \vspace{0.8cm}

    \begin{minipage}[t]{0.45\textwidth}
        \centering
        \resizebox{\linewidth}{!}{
            \begin{tikzpicture}
                \begin{yquantgroup}
                    \registers {
                        qubit {} q[4];
                    }
                    \circuit{
                        box {$R_{zz}(\theta_1)$} (q[0], q[3]);
                        box {$R_{zz}(\theta_2)$} (q[0], q[1]);
                        box {$R_{zz}(\theta_3)$} (q[1], q[2]);
                        box {$R_{zz}(\theta_4)$} (q[2], q[3]);
                    }
                \end{yquantgroup}
            \end{tikzpicture}
        }\\[6pt]
        \textbf{(c)} Shifted-circular alternating (SCA) entanglement layer
    \end{minipage}

    \vspace{0.8cm}

    \begin{minipage}[t]{0.45\textwidth}
        \centering
        \resizebox{\linewidth}{!}{
            \begin{tikzpicture}
                \begin{yquantgroup}
                    \registers {
                        qubit {} q[4];
                    }
                    \circuit{
                        box {$R_{zz}(\theta_1)$} (q[0], q[1]);
                        box {$R_{zz}(\theta_2)$} (q[0], q[2]);
                        box {$R_{zz}(\theta_3)$} (q[0], q[3]);
                        box {$R_{zz}(\theta_4)$} (q[1], q[2]);
                        box {$R_{zz}(\theta_5)$} (q[1], q[3]);
                        box {$R_{zz}(\theta_6)$} (q[2], q[3]);
                    }
                \end{yquantgroup}
            \end{tikzpicture}
        }\\[6pt]
        \textbf{(d)} Full entanglement layer
    \end{minipage}

    \caption{%
        This figure shows a single repetition of each circuit structures used in the experimental setup. For $n$ qubits, each structure is repeated $n$ times. All angles are randomly selected for each gate according to a uniform distribution \([- \pi, \pi]\). Each repetition is generated independently with independently-generated angles.
    }
    \label{fig:ExperimentalSetup}
\end{figure}

\subsection{SVD behavior}
The SVD threshold \(s_\text{max}\) plays a critical role in the performance of tensor network methods, as it directly controls the truncation of singular values, and thus, the growth of the bond dimensions of the intermediary MPO. Understanding its effect is essential before analyzing the scalability of the proposed MPO-based equivalence-checking method.

To investigate this, we generate 10 $5 \times 5$ circuits \(G\) of each entanglement type (linear, SCA, and full), and transpile them into new circuits \(G'\). We then check their equivalence using 60 logarithmically spaced SVD thresholds \(s_\text{max}\) in the range \([10^{-6}, 1]\).

\begin{figure}[t!]
    \centering
    \includegraphics[width=\linewidth]{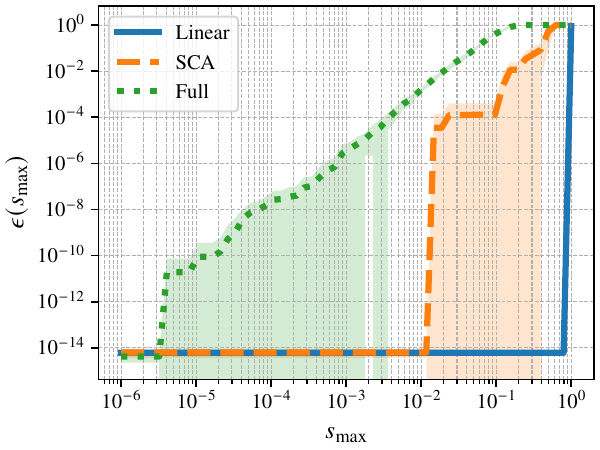}
    \caption{This figure shows the deviation $\epsilon(s_\text{max}) = 1 - \frac{\text{Tr}[W]}{2^n}$ caused by various SVD thresholds for each circuit type. This was tested for 60 logarithmically-spaced SVD thresholds \(s_\text{max}\) for 10 circuits of each type, where the average is plotted. The shaded area shows one standard deviation from the average.}
    \label{fig:SVDBehavior}
\end{figure}

For each \(s_\text{max}\), we compute the deviation from the ideal equivalence condition as:
\[
\epsilon(s_\text{max}) = 1 - \frac{\text{Tr}[W]}{2^n}.
\]
This \(\epsilon(s_\text{max})\) measures the inherent numerical error generated from the truncation in the method itself. This can be understood as the numerical tolerance required in the trace to assert equivalence between the two circuits \(G\) and \(G'\) for a given SVD threshold such that $\epsilon < \epsilon(s_\text{max})$ must be true to guarantee correctness. The average over 10 circuits of each entanglement type is plotted in \autoref{fig:SVDBehavior}.

The plot illustrates how \(\epsilon(s_\text{max})\) behaves for the three entanglement patterns, highlighting the threshold beyond which equivalence can no longer be reliably asserted. For instance, circuits with full entanglement exhibit significant sensitivity to the SVD threshold, continuously introducing more error as the threshold increases. In contrast, the SCA circuit with fewer long-range gates has a hard-cut from roughly machine error to a continuous curve around $s_\text{max} \approx 10^{-2}$. Notably, linear circuits are robust for any value $s_\text{max} < 1$. Due to the cancellations that occur in these benchmarks, the intermediary MPO will almost exclusively be close to the identity. As soon as the intermediary MPO is not consistently close, e.g., in the case of many long-range gates, then this value must be chosen with care.

\begin{figure*}[ht!]
    \centering
    \includegraphics[width=0.95\linewidth]{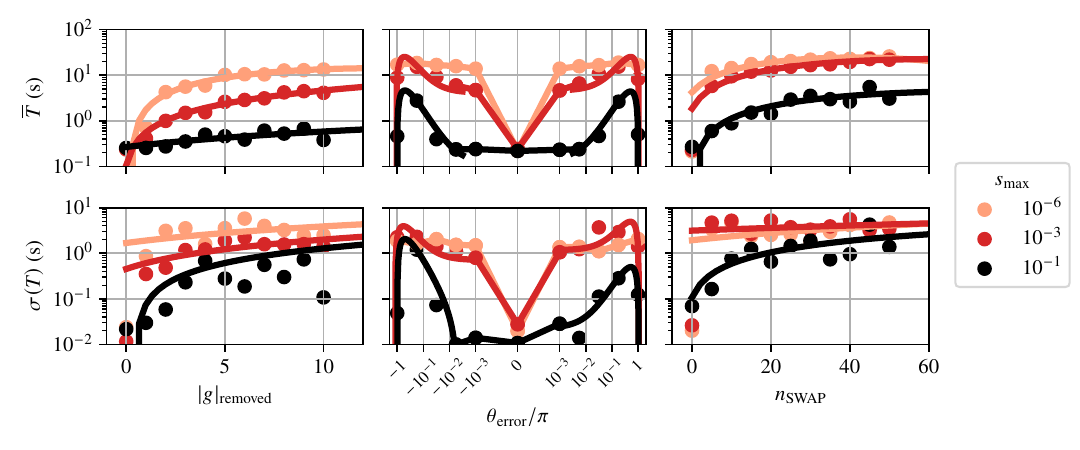}
    \caption{This figure compares different types and severity of errors on $n=10$ circuits with linear entanglement according to various SVD thresholds. Each color represents an SVD threshold for the MPO method. Each column corresponds to the error type such that the top row is the average runtime over 10 samples and the bottom is the data's standard deviation. The equivalent case can be identified at the zero-point of each plot.}
    \label{fig:SVDResults}
\end{figure*}

\subsection{Error injection} \label{sec:ErrorInjection}
 To investigate this, we tested three SVD thresholds, \(s_{\text{max}}\): a low threshold (\(10^{-6}\)), a medium threshold (\(10^{-3}\)), and a high threshold (\(10^{-1}\)). Since the equivalence condition \(G = G'\) holds only when the intermediary MPO exactly equals the identity and non-equivalent cases have infinite variations, we expect flexibility in choosing \(s_{\text{max}}\) without risking false equivalencies.

We conducted tests using linear entanglement circuits \(G\) of size \(n = 10\) (i.e., \(10 \times 10\)) and transpiled them into new circuits \(G'\), introducing various errors. We measured the runtime required to detect non-equivalence under different thresholds. Additionally, we evaluated the runtime in equivalent cases (no errors) to observe the effect of the SVD threshold on performance. These results are presented in \autoref{fig:SVDResults}.

\subsubsection{Missing gate errors}
In this test, we randomly removed a certain number of gates, \(|g|_\text{removed}\), from \(G'\) to simulate potential errors during transpilation. We observed that the runtime for each SVD threshold increases polynomially with the number of missing gates. As expected, the runtime decreases significantly with a higher threshold, and the standard deviation in runtime grows polynomially with the number of removed gates, albeit with multiple orders of magnitude difference in runtime. Interestingly, for low thresholds, the standard deviation is also roughly an order of magnitude smaller than the runtime. As \(|g|_\text{removed}\) increases, the standard deviation converges to similar values regardless of threshold, making the runtime more unpredictable.

In the equivalent case (\(|g|_\text{removed} = 0\)), the SVD threshold has negligible impact on the runtime. All thresholds yield similar results, with the lower threshold slightly reducing the standard deviation. This behavior likely results from the cancellation of gates during the spatial sweep, reducing the singular values to near zero in the equivalent case.

\subsubsection{Rotation angle errors}
In this test, we shifted all rotation angles in \(G'\) by a fixed value, \(\theta_{\text{error}}\), chosen from the range \([- \pi, \pi]\) using logarithmic scaling. The smallest error tested was \(\theta_{\text{error}} = 10^{-3} \pi\), with \(\theta_{\text{error}} = 0\) representing the equivalent case. The results reveal symmetry in runtime around the equivalent case, indicating that the sign of the error does not affect the performance. The standard deviation follows a similar trend, remaining about an order of magnitude smaller than the runtime.

For low and medium thresholds, any error causes a significant jump in runtime, reflecting a phase transition between the equivalent and non-equivalent cases. In contrast, the high threshold maintains a relatively constant runtime for small errors, likely because it disregards minor non-zero singular values. As \(\theta_{\text{error}}\) approaches \(\pi/4\), both the medium and high thresholds show an increase in runtime, likely due to the maximal operator entanglement in the intermediary MPO. For larger errors, the runtime decreases again as \(\theta_{\text{error}}\) approaches \(2\pi\), converging back to the equivalent case. The high threshold significantly reduces runtime, especially for small angle errors.

\subsubsection{Permutation errors}
In this test, we applied a random number of nearest-neighbor SWAP gates, \(n_{\text{SWAP}}\), at the beginning of \(G'\) to simulate mismatched virtual-physical qubits or incorrect qubit ordering. The runtime initially increases linearly with \(n_{\text{SWAP}}\) before stabilizing at a constant value as \(n_{\text{SWAP}} \rightarrow \infty\). The low and medium thresholds quickly reach this plateau with few SWAPs, while the high threshold scales more slowly with \(n_{\text{SWAP}}\) before also plateauing. The standard deviation follows a similar trend, with the high threshold showing higher deviation compared to the lower thresholds.
In the equivalent case, the low and medium thresholds are outliers, with significantly higher standard deviations, which stabilize after a single SWAP.

\begin{figure*}[ht!]
    \centering
    \includegraphics[width=\linewidth]{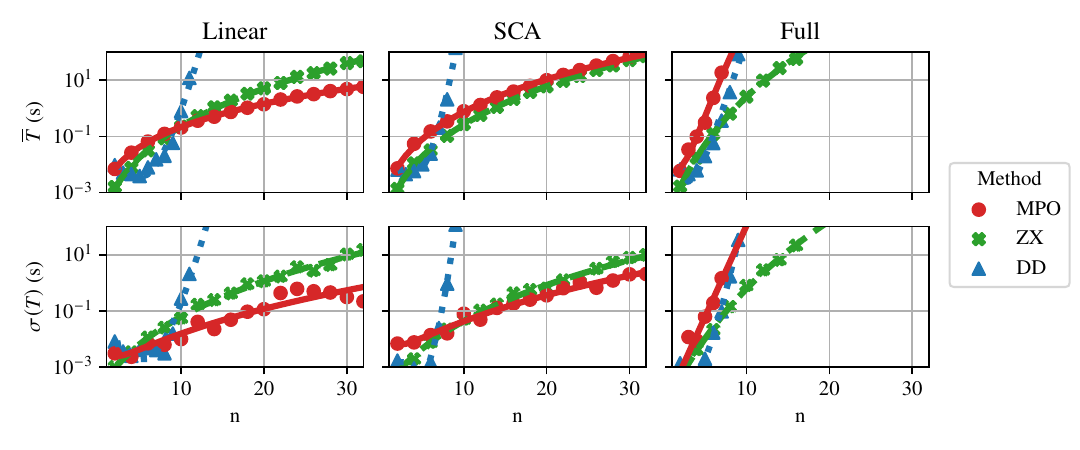}
    \caption{Equivalence scaling of each method with circuit size \(n\) across different entanglement patterns. Each color represents a different method, with the top row showing average runtime and the bottom row showing standard deviation over 10 samples.}
    \label{fig:EquivalenceResults}
\end{figure*}

\begin{figure*}[ht!]
    \centering
    \includegraphics[width=\linewidth]{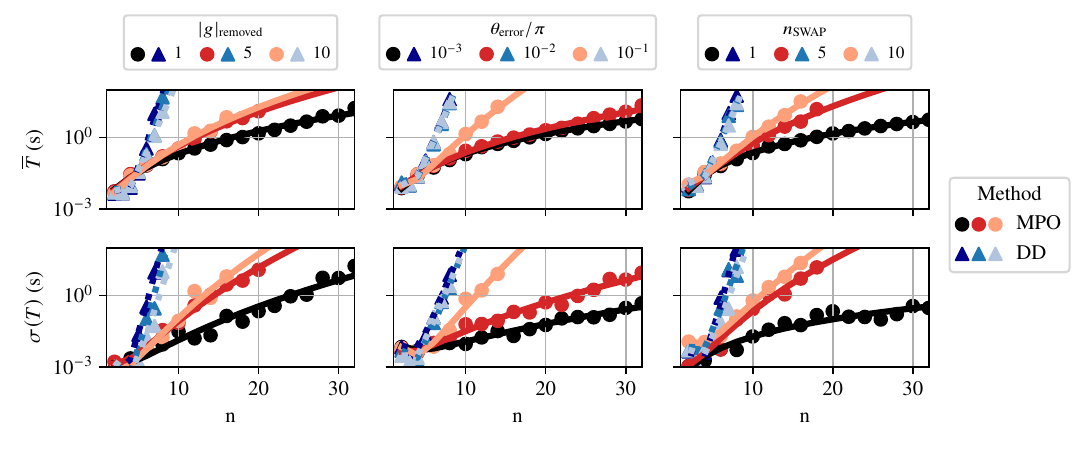}
    \caption{Non-equivalence scaling of each method with circuit size \(n\), considering different error types and severities applied to linear circuits. Each color represents a method, and each column corresponds to an error type. The top row shows the average runtime, and the bottom row shows the standard deviation over 10 samples.}
    \label{fig:NonEquivalenceResults}
\end{figure*}

\subsection{Scaling of equivalent circuits} \label{sec:EquivalenceScaling}
This section evaluates the scaling behavior of different methods for verifying the equivalence of circuits, repeated for increasing values of \(n\), to assess performance with circuit size. We compare the proposed MPO-based method against the DD- and ZX-based methods, with the results shown in \autoref{fig:EquivalenceResults}. All tests have an SVD threshold set such that we guarantee equivalence with a tolerance \(\epsilon = 10^{-13}\), i.e., linear $s_\text{max}=10^{-1}$, SCA $s_\text{max}=10^{-3}$, full $s_\text{max}=10^{-6}$.

For circuits with linear entanglement patterns, both the MPO-based and ZX methods scale polynomially with system size \(n\), whereas the DD method exhibits exponential scaling. The MPO-based method demonstrates better overall scaling than the ZX method for larger circuits.

As the number of long-range gates increases (i.e., in the SCA entanglement pattern), the MPO- and ZX-based methods continue to scale polynomially and perform comparably, while the DD method remains exponential. This indicates that for circuits with few long-range gates, the MPO-based method is competitive with other available methods.

For fully entangled circuits with many long-range gates, all methods approach exponential scaling, though the ZX method continues to scale more efficiently than the others. This suggests that an increase in long-range entanglement pushes all methods toward exponential growth in runtime, with the ZX method being the least affected by long-range gates. This is expected as the ZX calculus tends to be independent of the range of each gate, and depends more on the depth and number of operations.

Across all methods, the standard deviation in runtime scales exponentially, but it remains at least an order of magnitude lower than the average runtime. Notably, the MPO method shows exceptionally low standard deviation for the linear case, indicating stable and predictable performance as a function of \(n\).

\subsection{Scaling of non-equivalent circuits}
Here, we evaluate the scaling behavior for detecting non-equivalence between circuits. Since the ZX method cannot prove non-equivalence, we focus on comparing the MPO- and DD-based methods. Various errors of different intensity, as described in Sec. \ref{sec:ErrorInjection} and analyzed in Sec. \ref{sec:ErrorInjection}, were injected into linear entanglement circuits. For the MPO-based method, the high SVD threshold \(s_{\text{max}} = 10^{-1}\) was used to detect non-equivalence. The results are presented in \autoref{fig:NonEquivalenceResults}.

For all error types and severities, the DD method scales exponentially, likely because it becomes maximally large in all cases—an expected outcome given that it also scales exponentially for the equivalent case. In contrast, the MPO method scales polynomially, though more severe errors cause worse scaling, as expected due to the QMA-completeness of the equivalence-checking problem. These results can be extrapolated to understand the non-equivalence scaling when applied to SCA and full entanglement patterns, where we expect small errors to scale similarly to the equivalent case, and more severe to approach exponential scaling. Concretely, by combining the results in this section with the error plots in \autoref{sec:ErrorInjection}, we can estimate that the transition from polynomial to exponential scaling of runtime as a function of system size occurs approximately for phase errors \(\theta_{\text{error}} > 10^{-2} \pi\) or when the number of swapped gates reaches \(n_{\text{swap}} > 5\). While additional gate errors do worsen scaling behavior, the tests performed here do not exhibit true exponential scaling. We expect such scaling to emerge only for significantly more non-equivalent circuits as the removed gates increases \(|g|_\text{removed} \gg 10\).

Interestingly, for circuits with low-severity errors, the MPO method's scaling closely mirrors that of the linear equivalent case in Sec. \ref{sec:EquivalenceScaling}. This shows that the MPO method surpasses previous methods by far for detecting small degrees of non-equivalence.

The standard deviation scales exponentially, likely due to the wide variety of possible errors, though it remains more favorable for low-severity errors compared to the DD method. Additionally, the standard deviation grows in proportion to the average runtime, indicating that for very large circuits, the deviation could surpass the average runtime. However, at the current scale (\(n = 32\)), the MPO-based method remains stable and predictable, suggesting it is well-suited for near-term circuit development.

\section{Discussion} \label{sec:Discussion}
The MPO-based equivalence checking method introduced in this work demonstrates promising scalability for detecting both equivalent and non-equivalent circuits, marking an important advancement in the field of quantum circuit verification. Our approach shows superior scaling compared to DD-based methods, particularly for non-equivalence checking, and offers performance comparable to ZX-based methods for equivalent circuits. This makes the MPO-based method highly valuable as quantum circuits continue to grow in size and complexity.

One of the key strengths of the MPO-based approach is its ability to detect both minor and severe errors within quantum circuits, something that is crucial for a comprehensive equivalence checking tool. While the ZX method struggles with non-equivalence checking, the MPO method fills this gap, providing a more complete solution. However, it is important to acknowledge the inherent challenge in selecting a representative subset of quantum circuits and error types for evaluation. The vast diversity of quantum circuits and the multitude of possible errors make it difficult to cover all potential scenarios. The results presented here aim to provide a broad overview of the method's capabilities rather than an exhaustive evaluation.

Given that our method is based on tensor networks, several practical improvements could further enhance its performance. These include parallelizing the calculations \cite{huang_efficient_2021}, leveraging GPU acceleration \cite{lyakh_exatn_2022, bayraktar_cuquantum_2023}, and implementing a more optimized version than the current prototype. Long-range gates present the greatest bottleneck in scaling the method, as extending gates via identity tensors is not an ideal solution. One possible improvement could involve directly using gate tensors without extension, though this would introduce additional complexity in tracking dimensions. Alternatively, techniques from condensed matter physics, such as Krylov space methods for handling long-range interactions \cite{vanderstraeten_tangent-space_2019, paeckel_time-evolution_2019}, may offer more efficient solutions. Another potential optimization could be inspired by the identity-stripping technique used in DDs \cite{sander_stripping_2024}, which could analogously speed up the MPO method by removing unnecessary identity tensors.

Looking ahead, the MPO-based equivalence checking method opens several exciting avenues for future research and application. One notable direction is its potential use in quantum circuit optimization and compilation. By exploiting the method’s ability to detect non-equivalence, optimization techniques could be developed to iteratively refine quantum circuits until an equivalent, optimized version is generated. Additionally, the method could be applied to compare noisy, error-mitigated circuits with their ideal noiseless counterparts, offering a pathway toward improved error mitigation strategies in quantum computing.

In conclusion, the MPO-based method provides a scalable, flexible, and robust solution for quantum circuit equivalence checking. It addresses a key challenge in the field by offering a comprehensive approach to both equivalence and non-equivalence checking. With further development and optimization, this method holds significant potential for advancing quantum circuit verification, optimization, and error mitigation, paving the way for more reliable and scalable quantum computing systems.

\section{Acknowledgments}
This work received funding from the European Research Council (ERC) under the European Union’s Horizon 2020 research and innovation program (grant agreement No. 101001318) and Millenion,
grant agreement No. 101114305). This work was part of the Munich Quantum Valley, which is supported by
the Bavarian state government with funds from the Hightech Agenda Bayern Plus, and has been supported by the
BMWK on the basis of a decision by the German Bundestag through project QuaST, as well as by the BMK,
BMDW, and the State of Upper Austria in the frame of the COMET program (managed by the FFG).

\bibliography{library}

\end{document}